\documentclass[aps,showpacs,twocolumn,pra,superscriptaddress]{revtex4}
\usepackage{graphicx}

\usepackage{lipsum,mwe}

\usepackage{soul}

\usepackage{subfig}

\usepackage[T1]{fontenc}    
\usepackage[utf8]{inputenc} 
\usepackage{lmodern}        

\DeclareUnicodeCharacter{2212}{_}
\usepackage[utf8]{inputenc}
\usepackage{multirow}

\usepackage[tbtags]{amsmath}
\usepackage{amssymb}
\usepackage{amsfonts}
\usepackage{bmpsize}
\usepackage{mathtools}
\usepackage[colorlinks=true,citecolor=blue,urlcolor=blue]{hyperref}
\hypersetup{colorlinks,linkcolor=blue,filecolor=blue,urlcolor=blue,citecolor=blue}

\usepackage{times}

\usepackage{ragged2e}

\makeatletter
\def\@makecaption#1#2{%
  \vskip\abovecaptionskip
  \sbox\@tempboxa{\RaggedRight #1: #2}%
  \ifdim \wd\@tempboxa >\hsize
    \RaggedRight #1: #2\par
  \else
    \global \@minipagefalse
    \hb@xt@\hsize{\hfil\box\@tempboxa\hfil}%
  \fi
  \vskip\belowcaptionskip}
\makeatother

\begin{document}

\title{
Complementarity Reveals Entanglement Sharing in Sequential Quantum Measurements}

\author{Zinuo Cai}
\affiliation{Key Laboratory of Low-Dimension Quantum Structures and Quantum Control of Ministry of Education, Synergetic Innovation Center for Quantum Effects and Applications, Xiangjiang-Laboratory and Department of Physics, Hunan Normal University, Changsha 410081, China}
    \affiliation{Hunan Research Center of the Basic Discipline for Quantum Effects and Quantum Technologies, Hunan Normal University, Changsha 410081, China}
	\affiliation{Institute of Interdisciplinary Studies, Hunan Normal University, Changsha 410081, China}

\author{Changliang Ren}\thanks{renchangliang@hunnu.edu.cn}
\affiliation{Key Laboratory of Low-Dimension Quantum Structures and Quantum Control of Ministry of Education, Synergetic Innovation Center for Quantum Effects and Applications, Xiangjiang-Laboratory and Department of Physics, Hunan Normal University, Changsha 410081, China}
    \affiliation{Hunan Research Center of the Basic Discipline for Quantum Effects and Quantum Technologies, Hunan Normal University, Changsha 410081, China}
	\affiliation{Institute of Interdisciplinary Studies, Hunan Normal University, Changsha 410081, China}

\begin{abstract}

{We investigate entanglement sharing in a two-qubit sequential measurement scenario using three complementary classical correlation metrics: mutual information ($\mathcal{I}$), sum of conditional probabilities ($\mathcal{S}$), and the Pearson correlation coefficient ($\mathcal{C}$). By investigating both weak measurement and probabilistic projective measurement (PPM) strategies in unilateral and bilateral scenarios, the phenomenon of entanglement sharing is conclusively certified when multiple pairs of classical correlation metrics simultaneously exceed their thresholds. Our investigation reveals that weak measurement strategies are more favorable than PPM for exhibiting entanglement sharing, regardless of the scenario. Furthermore, the mutual information criterion fails to characterize entanglement sharing in the bilateral scenario. While, the Pearson correlation criterion ($\mathcal{C}$) is proven to be the most robust across all strategies and scenarios. These findings unveil a critical trade-off between measurement disturbance and complementary correlation recovery, which is essential for quantum resource reuse problems.
}

\end{abstract}


\maketitle

\section{Introduction}







{\color{black}

Quantum systems can exhibit correlations that cannot be simulated by any classical model, which lies at the heart of quantum theory. Among these nonclassical correlations, quantum entanglement plays a central role. It is typically identified through criteria such as the positive partial transpose (PPT) criteria \cite{Peres_PhysRevLett.77.1413_1996}, entanglement witnesses \cite{horodecki2009quantum,Lewenstein_PhysRevA.62.052310_2000,Brand_PhysRevA.72.022310_2005,Kiesel_PhysRevLett.95.210502_2005}, entropy-based criteria \cite{Horodecki_pla147_1994,Horodecki+pla377_1996,Cerf_PhysRevLett.79.5194_1997}, or covariance-matrix approaches\cite{Devi_PhysRevLett.98.060501_2007,Duan_PhysRevLett.84.2722_2000,
Giedke_PhysRevLett.87.167904_2001}. The essential aim of these methods is to characterize the inseparability of complex quantum systems.
Recent studies, however, have demonstrated that the principle of complementarity can also serve as an effective tool for detecting and quantifying entanglement \cite{Bennett_PhysRevLett.70.1895_1993,Maccone_Complementarity_2015}. In contrast to those traditional approaches, complementarity offers a more operational perspective: it reveals how statistical correlations between measurements in different bases reflect the nonclassical structure of a quantum state, thereby formally integrating the quantum system, the measurement process, and the choice of observables into a unified framework.


Mathematically, the principle of complementarity can be formalized by pairs of mutually unbiased bases (MUBs)~\cite{Schwinger_UNITARYOB_1960,Ivonovic_1981_Geometrical,kraus_complementary_1987,William_Optimal_1989}. Such as, in a $d$-dimensional Hilbert space, two bases ${|a_i\rangle}$ and ${|c_j\rangle}$ are mutually unbiased when $|\langle a_i | c_j \rangle|^2 = d^{-1}$ for all $i, j$. The maximum number of mutually unbiased bases is bounded by $d+1$. A profound implication of this structure is that any violation of the classical correlation bound within a pair of MUBs furnishes a criterion for entanglement detection~\cite{Huang_PhysRevA.82.012335_2010,Spengler_PhysRevA.86.022311_2012,Maccone_PhysRevLett.114.130401_2015,Paul_PhysRevA.94.012303_2016}. Furthermore, when we extend this idea to investigate sequential measurement scenarios involving multiple parties, may reveal that complementarity can not only detect the presence of entanglement but also elucidate how it is shared among the participants.


In fact, the problem of sharing quantum correlations triggered by quantum measurements has attracted considerable attention over the past decade~\cite{cairen_2024_review}. These investigations are not only of fundamental theoretical interest but also raise novel questions concerning the recycling and reuse of quantum resources. Since the pioneering work by \citet{Silva.Ralph_PhysRevLett.114.250401_2015} in 2015, which demonstrated the sharing of nonlocality among multiple observers via weak measurement strategies, this field has developed rapidly, including active and passive non-local sharing based on different motivations \cite{Ren.Changliang_PhysRevA.100.052121_2019}, asymmetric positive operator-valued measure(POVM) strategies~\cite{Brown.Peter.J_PhysRevLett.125.090401_2020}, sequential projective measurement strategies~\cite{Anna._PhysRevLett.129.230402_2022,Dong_PhysRevA.110.012203_2024,Sasmal_2023_arXiv_unbounded}, as well as scenarios involving two, three, and even multiple parties~\cite{YaoDan.PhysRevA.103.052207_2021,Ren.Changliang_PhysRevA.105.052221_2022,Zhu.Jie_PhysRevA.105.032211_2022,Hou_PhysRevA.105.042436_2022,Cai_jpafull_2024}.
Several experimental verifications have also been reported~\cite{Schiavon._Quantum.Sci.Technol.2_015010_2017,
Hu.Meng.Jun_Quantuminf.s41534-018-0115-x_2018,
Feng.Tianfeng_PhysRevA.102.032220_2020,
Foletto.Giulio_PHYS.REV.APPL.13.044008_2020}, gradually forming a systematic theoretical and experimental framework. In contrast, the sharing of entanglement remains comparatively less explored. Although several works have shown that entanglement can be detected by an arbitrary number of observers through entanglement witnesses under sequential single- or two-sided measurements~\cite{
Srivastava.Chirag_PhysRevA.105.062413_2022,
Pandit.Mahasweta_PhysRevA.106.032419_2022,
HuMingLiang_PhysRevA.108.012423},
the overall systematic understanding remains less developed than that of nonlocality sharing. In particular, the role of complementarity-based entanglement criteria in the context of entanglement sharing has yet to be fully analyzed and explored.

}

Complementarity-based criteria, in our view, not only provide an alternative approach for entanglement detection but also have the potential to provide new research perspectives for operational quantum entanglement sharing. 
Motivated by the above insights, we systematically investigate the classical correlations of complementary measurement outcomes in sequential scenarios. For both one-sided and two-sided sequential measurements, and within weak-measurement and probabilistic projective measurement (PPM) strategies, we evaluate three complementary correlation measures—mutual information, the sum of conditional probabilities, and the Pearson correlation coefficient. We show how these measures capture correlations among different observer combinations. Crucially, when the correlations of distinct pairs simultaneously exceed their respective thresholds, the presence of entanglement sharing is conclusively established. For each strategy, we further contrast symmetric and asymmetric measurement configurations. Our results show that the Pearson correlation coefficient consistently serves as the most effective criterion. Compared with PPM, weak measurements can more easier exhibit the phenomenon entanglement sharing, as they disturb the state less. 
{\color{black}
Under the weak measurement strategy, the mutual information criterion performs better in asymmetric configurations of the unilateral scenario. However, in the bilateral scenario, it fails to capture entanglement sharing. This contrasts with the other two criteria, which perform better under symmetric configurations and are applicable to both unilateral and bilateral scenarios. Under the PPM strategy, the optimal configurations for the criteria are largely the same as those under weak measurements, with one exception: for the sum of conditional probabilities criterion in the bilateral scenario, the optimal configuration becomes asymmetric.
}

{\color{black}
The paper is structured as follows. We begin by introducing the theoretical framework, including the model, three complementarity criteria, and two measurement strategies. In Section~\ref{onside}, we provide a detailed analysis of complementary correlations and their role in entanglement sharing within a unilateral sequential scenario. Section~\ref{twoside} extends the discussion to a bilateral sequential scenario. we conclude in  Section~\ref{Con} with a summary.
}

\section{Theoretical Model, Complementarity Criteria, and Measurement Strategies}\label{scenario}

In this section, we first introduce the quantum system model under investigation along with its key characteristics. Our focus is on a two-qubit system where we explore the phenomenon of entanglement sharing. To characterize the complementarity properties of the quantum states involved, we employ three correlation criteria based on complementarity. These criteria effectively capture the distinctions between quantum and classical correlations, providing reliable means to detect both the presence and sharing of entanglement. Regarding the measurement process, we consider two representative strategies: weak measurements and probabilistic projective measurements. These approaches reflect different physical aspects, namely measurement strength and randomness. By comparing the effects of these strategies on the structure of entanglement sharing within the system, we reveal the crucial role of measurement choice in the distribution of quantum resources. We describe how the marginal probabilities are derived from the joint probabilities obtained through sequential measurements. These marginal probabilities are subsequently employed to investigate correlation sharing.

\subsection{Theoretical Model}

\begin{figure*}[htbp]
	\centering
	\includegraphics[width=0.7\textwidth]{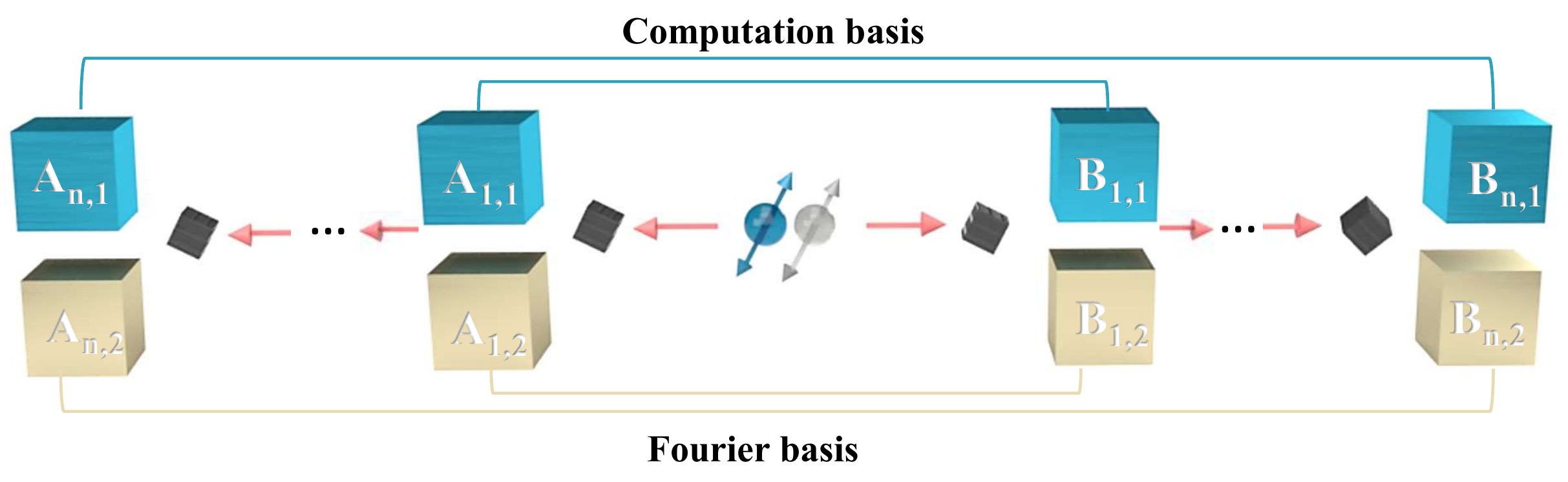}
	\caption{\small{The general bilateral sequential scenario: Schematic of a quantum scenario where a pair of qubits emit particles to sequential observers on both sides. Each observer performs measurements with two complementary choices: for each $k\in(1,2,...,n)$, Alice$_k$ measures with observables $\hat{A}_{k,1}$ or $\hat{A}_{k,2}$, and Bob$_k$ with $\hat{B}_{k,1}$ or $\hat{B}_{k,2}$. Here, $A_{k,1}$ and $B_{k,1}$ are defined in the computational basis, while $A_{k,2}$ and $B_{k,2}$ are defined in the Fourier basis. Arrows indicate the transfer of qubits. The black dice symbolize that each observer's measurement choice is made probabilistically.}}
	\label{general}
\end{figure*}

As illustrated in Fig. \ref{general}, we consider the standard Bell nonlocality sharing scenario, where a source generates a two-qubit state distributed to two sets of observers, $\{\mathrm{Alice_n}\}$ and $\{\mathrm{Bob_n}\}$ respectively. Without loss of generality, we assume that they share the following bipartite state, 
\begin{equation}
    |\psi\rangle=\cos\theta|00\rangle+\sin\theta|11\rangle
\end{equation}
where the density matrix is $\rho=|\psi\rangle\langle\psi|$.
Each observer $\mathrm{Alice_n}$ ($\mathrm{Bob_n}$) chooses two different dichotomic observables independently, denoted by $\hat{A}_{n,m}$ ($\{\hat{B}_{n,m}\}$), with outcomes $\{a_n\}$ ($\{b_n\}$), where $m$ denotes two distinct measurement choices $m\in\{1,2\}$.
Our goal is to analyze the correlation properties of the quantum system based on the measurement outcomes of these observers, and to imply how entanglement is sequentially shared by these observers.



\subsection{ Complementarity Criteria}

This work aims to interpret quantum entanglement in terms of classical correlations arising from measurements of complementary observables. In previous studies~\cite{Maccone_PhysRevLett.114.130401_2015}, quantum correlations were revealed through complementary measurements in scenarios involving only a single observer on each side—that is, a minimal $\text{Alice}_1$-$\text{Bob}_1$. As a starting point, we introduce the complementarity-based correlation criteria employed in that setting. Consider a bipartite entangled state distributed to two observers, each performing measurements in a set of mutually complementary bases. Without loss of generality, we define the measurement operators $\{\hat{A}_{1,1},\hat{B}_{1,1}\}$ to correspond to the computational basis \( \{ |0\rangle, |1\rangle \} \), while  $\{\hat{A}_{1,2},\hat{B}_{1,2}\}$ are defined in the conjugate Fourier basis \( \left\{ \frac{|0\rangle + |1\rangle}{\sqrt{2}}, \frac{|0\rangle - |1\rangle}{\sqrt{2}} \right\} \). Since \( n = 1 \) in this case, for simplicity, we omit the subscript \( n \) in the notation of measurement operators throughout this section.

To quantify the correlations between measurement outcomes, we introduce correlation measures,  $\mathcal{M}_{A_{1}B_{1}}$ and $\mathcal{M}_{A_{2}B_{2}}$, corresponding to the observable pairs $\{\hat{A}_{1,1},\hat{B}_{1,1}\}$ and $\{\hat{A}_{1,2},\hat{B}_{1,2}\}$, respectively. Here, we have identified three specific measures to quantify these correlations: mutual information $\mathcal{M}_{XY}=\mathcal{I}_{XY}$, the sum of conditional probabilities $\mathcal{M}_{XY}=\mathcal{S}_{XY}$, the Pearson correlation coefficient $\mathcal{M}_{XY}=\mathcal{C}_{XY}$. The complementary correlations can be given as the sum of the absolute value of the two measures $|\mathcal{M}_{A_{1}B_{1}}+\mathcal{M}_{A_{2}B_{2}}|$.

\textit{Mutual information}——The first measure is the mutual information: $\mathcal{I}_{A_{1}B_{1}}=H(A_{1})-H(A_{1}|B_{1})$, where $H(A_{1})$ represents the Shannon entropy of the measurement outcome probabilities for the first system, while $H(A_{1}|B_{1})$ denotes the conditional entropy of the first system's outcomes conditioned on the second system. The definition are $H(A_{1})=-p(a|\hat{A}_{1})\log_2 p(a|\hat{A}_{1})$, and $H(A_{1}|B_{1})=-\sum_{a,b}
p(a|b;\hat{A}_{1},\hat{B}_{1}) p(b|\hat{B}_{1})\log_2 p(a|b;\hat{A}_{1},\hat{B}_{1})$,
where \( p(a\,|\,b,\, \hat{A}_1\,|\,\hat{B}_1) \) denotes the conditional probability of obtaining outcome \( a \) on the first subsystem, given that outcome \( b \) is observed on the second subsystem under the joint measurement setting \( \hat{A}_1 \) and \(\hat{B}_1 \).
 $\mathcal{I}_{A_{2}B_{2}}$ can be defined similarly.
In this scenario, the bipartite quantum state is maximally entangled if and only if $\mathcal{I}_{A_{1}B_{1}}+\mathcal{I}_{A_{2}B_{2}}=2\log_2 d$. Furthermore, if $\mathcal{I}_{A_{1}B_{1}}+\mathcal{I}_{A_{2}B_{2}}>\log_2 d$, the bipartite system can be deemed entangled.

\textit{Sum of conditional probabilities}—The second measure is the sum of conditional probabilities:
\begin{align}\label{sab}
    \mathcal{S}_{A_{1}B_{1}}=\sum_{a_1,b_1}p(a_1|b_1;\hat{A}_{1},\hat{B}_{1}).
\end{align}
And $\mathcal{S}_{A_{2}B_{2}}$ can be defined similarly.
In this scenario, the bipartite quantum state is maximally entangled if and only if there exist two complementary bases such that $\mathcal{S}_{A_{1}B_{1}}+\mathcal{S}_{A_{2}B_{2}}=2d$. If $\mathcal{S}_{A_{1}B_{1}}+\mathcal{S}_{A_{2}B_{2}}>3$ or $\mathcal{S}_{A_{1}B_{1}}+\mathcal{S}_{A_{2}B_{2}}<1$, the bipartite system is entangled.

\textit{Pearson corrlation}——The third measure is the Pearson correlation coeffcient: 
\begin{align}
\mathcal{C}_{A_{1}B_{1}}=\frac{\langle A_{1}B_{1} \rangle- \langle A_{1}\rangle \langle B_{1}\rangle }{\Delta_{A_{1}}\Delta_{B_{1}}}, \end{align}
where $\langle X\rangle=\mathrm{Tr}[\hat{X}\rho]$ is the expectation value on the quantum state $\rho$. $\Delta_X^2$ is the variance of $\hat{X}$.
In this scenario, the state is maximally entangled if and only if $|\mathcal{C}_{A_{1}B_{1}}|+|\mathcal{C}_{A_{2}B_{2}}|=2$. If $|\mathcal{C}_{A_{1}B_{1}}|+|\mathcal{C}_{A_{2}B_{2}}|>1$, the two systems are entangled. 


\subsection{ Sharing Quantum Correlations: Strategies and Measurements}

In ~\cite{Maccone_PhysRevLett.114.130401_2015}, both \( \mathrm{Alice}_1 \) and \( \mathrm{Bob}_1 \) perform strong projective measurements, which completely destroy the correlations between the two qubits. As a result, no nonclassical features can be observed in the measurements of subsequent observers. To explore the sharing of quantum correlations via complementary measurements in a sequential measurement scenario, it is essential that intermediate observers employ nonprojective measurements in mutually unbiased bases. Intermediate observers ($1 \leq k \leq n-1$) employ quantum non-destructive measurements to preserve state correlations, while terminal observers ($\text{Alice}_n$ and $\text{Bob}_n$) perform strong measurements. 

For any aribitry pair of observers, \( \mathrm{Alice}_k \) and \( \mathrm{Bob}_k \) independently choose between two binary-outcome measurement operators. As defined earlier, the measurement observables $\{\hat{A}_{k,1},\hat{A}_{k,2}\}$ correspond to \( \mathrm{Alice}_k \), while $\{\hat{B}_{k,1},\hat{B}_{k,2}\}$ correspond to \( \mathrm{Bob}_k \). The observables $\hat{A}_{k,1}$ and $\hat{B}_{k,1}$  are constructed from one common complete orthonormal basis $\{|\mu_i\rangle\}$, with $\hat{A}_{k,1}=\sum_i f_{\hat{A}_{k,1}}(\mu_i)\,|\mu_i\rangle\langle\mu_i|$ and $\hat{B}_{k,1}=\sum_i f_{\hat{B}_{k,1}}(\mu_i)\,|\mu_i\rangle\langle\mu_i|$, respectively. Whereas $\hat{A}_{k,2}$ and $\hat{B}_{k,2}$ are defined with respect to another mutually unbiased basis $\{|\nu_j\rangle\}$, with $\hat{A}_{k,2}=\sum_j g_{\hat{A}_{k,2}}(\nu_j)\,|\nu_j\rangle\langle\nu_j|$ and $\hat{B}_{k,2}=\sum_j g_{\hat{B}_{k,2}}(\nu_j)\,|\nu_j\rangle\langle\nu_j|$, respectively. So $\{\hat{A}_{k,2},\hat{A}_{k,2}\}$ (or $\{\hat{B}_{k,1},\hat{B}_{k,2}\}$) are complementary
properties as they satisfy $|\langle \mu_i | \nu_j \rangle|^2 = \frac{1}{d}$ for all $i,j$ in $d$-dimensional Hilbert space.
Within the general framework, these operators can be represented as positive operator-valued measures (POVMs).

In this study, the realization of quantum correlation sharing requires a careful trade-off between information gain and system disturbance inherent in non-destructive measurements, making the choice of measurement strategy crucial. In such scenarios, measurements not only extract information but also modify the quantum state through post-selection. Different types of non-destructive measurements lead to distinct patterns of correlation redistribution.

Specifically, in our discussion, the non-intermediate observers, \( \text{Alice}_n \) and \( \text{Bob}_n \), perform strong projective measurements in the \( \{ \hat{\sigma}_x, \hat{\sigma}_z \} \) bases. All intermediate observers, \( \text{Alice}_k \) and \( \text{Bob}_k \) (\( 1 \leq k \leq n{-}1 \)), employ one of two measurement strategies: the weak measurement scheme or the probabilistic projective measurement (PPM) scheme. These approaches balance the trade-off between measurement invasiveness and experimental feasibility, including potential noise effects, and thus enable a comprehensive analysis of entanglement sharing in sequential measurement scenarios.

The first strategy involves implementing complementary measurements via a weak measurement process by intermediate observers. In this approach, the incoming particle is measured with weak strength, inducing only minimal disturbance to its quantum state. As a result, a significant portion of the original quantum information is preserved, enabling the partial retention of quantum correlations for subsequent measurements. The weak measurement operators are defined as $\hat{A}_{k,1}=\eta_{ _{\hat{A}_{k,1}}}\hat{\sigma}_z$, $\hat{A}_{k,2}=\eta_{\hat{A}_{k,2}}\hat{\sigma}_x$, where $0<\eta_{i}\le1$ denotes the measurement strength. $\hat{B}_{k,1}$ and $\hat{B}_{k,2}$ can be defined similarly. The eigenstates corresponding to these operators are $\{|\mu_{\pm}\rangle\}\in\{|0\rangle,|1\rangle\}$ for $\hat{A}_{k,1}$ and $\{|\nu_{\pm}\rangle\}\in\{\frac{|0\rangle+|1\rangle}{\sqrt{2}},\frac{|0\rangle-|1\rangle}{\sqrt{2}}\}$ for $\hat{A}_{k,2}$. The orthogonality condition $|\langle \mu_{\pm}|\nu_{\pm}\rangle|^2=\frac{1}{d}$ is satisfied. Therefore, the weak measurement strategy can be utilized for the exploration of complementary correlations.
These operators can be constructed using two Kraus operators. For $\hat{A}_{k,1}$, they are  $\mathcal{K}^{\hat{A}_{k,1}}_1=\sqrt{\frac{1+\eta_{\hat{A}_{k,1}}}{2}}|\mu_{+}\rangle\langle\mu_{+}|+\sqrt{\frac{1-\eta_{\hat{A}_{k,1}}}{2}}|\mu_{-}\rangle\langle\mu_{-}|$ and  $\mathcal{K}^{\hat{A}_{k,1}}_2=\sqrt{\frac{1-\eta_{\hat{A}_{k,1}}}{2}}|\mu_{+}\rangle\langle\mu_{+}|+\sqrt{\frac{1+\eta_{\hat{A}_{k,1}}}{2}}|\mu_{-}\rangle\langle\mu_{-}|$. The effective POVM measurements are then given by $E_{\hat{A}_{k,1}}^+(\eta_{\hat{A}_{k,1}})=(\mathcal{K}^{\hat{A}_{k,1}}_1)^\dagger(\mathcal{K}^{\hat{A}_{k,1}}_1)$, $E_{\hat{A}_{k,1}}^-(\eta_{\hat{A}_{k,1}})=(\mathcal{K}^{\hat{A}_{k,1}}_2)^\dagger(\mathcal{K}^{\hat{A}_{k,1}}_2)$. Consequently, they can be expressed as $E_{\hat{A}_{k,1}}^+(\eta_{\hat{A}_{k,1}})=\frac{1+\eta_{\hat{A}_{k,1}}}{2}|\mu_{+}\rangle\langle\mu_{+}|+\frac{1-\eta_{\hat{A}_{k,1}}}{2}|\mu_{-}\rangle\langle\mu_{-}|$, 
$E_{\hat{A}_{k,1}}^-(\eta_{\hat{A}_{k,1}})=\frac{1-\eta_{\hat{A}_{k,1}}}{2}|\mu_{+}\rangle\langle\mu_{+}|+\frac{1+\eta_{\hat{A}_{k,1}}}{2}|\mu_{-}\rangle\langle\mu_{-}|$,
which correspond to the measurement outcomes of ``$+1$" and ``$-1$" respectively. The same applies to the other three measurement operators. For this type of measurement, when interpreted in terms of information gain and state disturbance, the measurement strength $\eta$ is equivalent to the information-gain factor  $G$  defined in Ref.~\cite{Silva.Ralph_PhysRevLett.114.250401_2015}, while the disturbance factor $F$, characterizing how well the system remains undisturbed, has been shown to satisfy $F=\sqrt{1-\eta^2}$~\cite{cairen_2024_review} within this framework.


The second measurement strategy is PPM. Assumed that any arbitrary intermediate observer with a coin that has a probability $\alpha$ of landing heads and a probability $1-\alpha$ of tails. Each observer determines their measurement strategy based on the outcome of a local coin toss. For instance, the two POVM measurements employed by \( \mathrm{Alice}_k \) can be represented as follows, $\hat{A}_{k,1}=\alpha_{\hat{A}_{k,1}} \sigma_z+(1-\alpha_{\hat{A}_{k,1}}) \mathbb{I},\hat{A}_{k,2}=\alpha_{\hat{A}_{k,2}} \sigma_x+(1-\alpha_{\hat{A}_{k,2}}) \mathbb{I}$, where $0<\alpha_{i}\le1$ represents the probability weight. This measurement can be implemented as a POVM constructed from three Kraus operators. For the observable \( \hat{A}_{k,1} \), the corresponding Kraus operators are given by \(\mathcal{K}^{\hat{A}_{k,1}}_1 = \sqrt{\alpha_{\hat{A}_{k,1}}} |\mu_{+}\rangle\langle \mu_{+}|, \quad
\mathcal{K}^{\hat{A}_{k,1}}_2 = \sqrt{\alpha_{\hat{A}_{k,1}}} |\mu_{-}\rangle\langle \mu_{-}|, \quad
\mathcal{K}^{\hat{A}_{k,1}}_3 = \sqrt{1 - \alpha_{\hat{A}_{k,1}}} \, \mathbb{I},
\)
where \( 0 < \alpha_{\hat{A}_{k,1}} \leq 1 \) denotes the measurement strength and \( \{ |\mu_{\pm}\rangle \} \) are orthonormal basis vectors. By collecting measurement statistics through classical post-processing of projective outcomes, each intermediate observer effectively performs a two-element POVM characterized by the following positive operators, \(
F_{\hat{A}_{k,1}}^{+} = (\mathcal{K}^{\hat{A}_{k,1}}_1)^{\dagger} \mathcal{K}^{\hat{A}_{k,1}}_1 + (\mathcal{K}^{\hat{A}_{k,1}}_3)^{\dagger} \mathcal{K}^{\hat{A}_{k,1}}_3 = \alpha_{\hat{A}_{k,1}} |\mu_{+}\rangle\langle \mu_{+}| + (1 - \alpha_{\hat{A}_{k,1}}) \mathbb{I},
\)
\(F_{\hat{A}_{k,1}}^{-} = (\mathcal{K}^{\hat{A}_{k,1}}_2)^{\dagger} \mathcal{K}^{\hat{A}_{k,1}}_2 = \alpha_{\hat{A}_{k,1}} |\mu_{-}\rangle\langle \mu_{-}|,
\) corresponding to the measurement outcomes ``\(+1\)'' and ``\(-1\)'', respectively. The same applies to
the other three measurement operators. In this measurement process, the probability $\alpha$ is equivalent to the information-gain factor $G$ defined in Ref.~\cite{Silva.Ralph_PhysRevLett.114.250401_2015}, while the disturbance factor $F$, which characterizes the extent to which the system remains undisturbed, is clearly given by $(1-\alpha)$.

Crucially, these two measurement strategies exhibit fundamentally distinct operational natures. In the case of weak measurements, the sharpness parameter \( \eta \) originates from the intrinsic quantum indistinguishability of the measurement device's internal states. In contrast, the parameter \( \alpha \) characterizing probabilistic projective measurements (PPMs) arises purely from classical local randomness, with no quantum mechanical origin. Although these two physically distinct procedures may yield identical outcome probability distributions, they induce fundamentally different post-measurement states in the overall system due to their differing mechanisms of state disturbance and post-selection.

\subsection{Measurement Process and the marginal probabilities}

In the scenario described above, supposed that the measurement operator associated with an arbitrary \( k \)-th ($k\in\{1,...n\}$) pair of observers, $\hat{A}_{k,m}$  and $\hat{B}_{k,m}$, can then be decomposed as $\{\hat{A}_{k,m}^+,\hat{A}_{k,m}^-\}=\{\hat{A}_{k,m}^{a_k}\}$ and $\{\hat{B}_{k,m}^+,\hat{B}_{k,m}^-\}=\{\hat{B}_{k,m}^{b_k}\}$, where ``$\pm$'' correspond to the measurement outcomes $\{a_k,b_k\}\in\{+1,-1\}$. When the initial state is $\rho$, the post-measurement state after the first pair of observers performs their measurements, can be written as
\begin{align}
	\rho_{1} = (\hat{A}_{1,m}^{a_1} \otimes \hat{B}_{1,m}^{b_1}). \rho. (\hat{A}_{1,m}^{a_1} \otimes \hat{B}_{1,m}^{b_1})^\dagger,
\end{align}
where the corresponding measurement results are $a_1$ and $b_1$ respectively. After $n$ pairs of observers’ sequential measurements, the quantum state can be expressed as
\begin{align}
	\rho_{n} = (\hat{A}_{n,m}^{a_n} \otimes \hat{B}_{n,m}^{b_n}) .\rho_{n-1}. (\hat{A}_{n,m}^{a_n} \otimes \hat{B}_{n,m}^{b_n})^\dagger
\end{align}
where the corresponding measurement results for $\{\mathrm{Alice_k}\}$ and $\{\mathrm{Bob_k}\}$ are $a_k$ and $b_k$ respectively. Therefore, the joint probability distributions can be easily obtained as
\begin{small}
\begin{align}
	P(a_1...a_n,b_1.. b_n | \hat{A}_{1,m}... \hat{A}_{n,m}, \hat{B}_{1,m}... \hat{B}_{n,m}) = \mathrm{Tr}[\rho_n].
\end{align}
\end{small}
For any arbitrary $k$-th pair of observers, the marginal measurement probability can be obtained,
\begin{small}
\begin{align}\label{joint measurement result-1}
	P(a_k,b_k|\hat{A}_{k,m},\hat{B}_{k,m})
		\nonumber\\=\sum_{k'\neq k} P(a_1...a_n,& b_1.. b_n | \hat{A}_{1,m}... \hat{A}_{n,m}, \hat{B}_{1,m}... \hat{B}_{n,m}).
\end{align}
\end{small}
Based on the obtained probability distributions, the corresponding Shannon entropies can be given, which in turn allow the evaluation of correlation features via complementarity-based criteria.

\section{Unilateral Sequential Entanglement Sharing}\label{onside}

{\color{black}
We start with the unilateral sequential scenario, where sequential measurements are performed on a single particle from one side. In this scenario, we investigate the entanglement between Alice$_1$ and Bob$_1$ as well as between Alice$_2$ and Bob$_1$, where distinguish the different correlations by $k\in\{1,2\}$. The corresponding measurements of Alice$_k$ (Bob$_{k'}$) are denoted by $\hat{A}_{k,m}$ ($\hat{B}_{k',m}$), where, for simplicity, the indices $k$ ($k'$) will be omitted. To quantify these correlations, we employ three complementary criteria,  $\{\mathcal{I}^{(k)}_{A_1B_1}+\mathcal{I}^{(k)}_{A_2B_2},\mathcal{S}^{(k)}_{A_1B_1}+\mathcal{S}^{(k)}_{A_2B_2},\mathcal{C}^{(k)}_{A_1B_1}+\mathcal{C}^{(k)}_{A_2B_2}\}$, with $k\in\{1,2\}$, all of which can be constructed from the joint measurement probability distribution as defined in Eq.~(\ref{joint measurement result-1}). 
}

\subsection{Unilateral Sharing Via Weak Measurement Strategy}

We first adopt the weak measurement strategy in this scenario. Without loss of generality, we assume that the initial state is an arbitrary two-qubit state and analyze the sharing of quantum correlations using three complementary criteria. 

For the mutual information criterion, we note that $H^{(k)}(A_1)\leq \log_2[d]$ and $H^{(k)}(A_2)\leq \log_2[d]$. Since the information gain parameter $G_i=\eta_{\hat{A}_{1,i}}$~($i \in \{1, 2\}$), and the degree to which the system remains undisturbed is represented by the factor $F_i =\sqrt{1 - \eta_i^2}= \sqrt{1-G_i^2}$, substituting it into the joint probability distribution, 
{\color{black}
we can obtain the entropies $H^{(1)}(A_1)$, $H^{(2)}(A_1)$, $H^{(k)}(A_1|B_1)$ and $H^{(k)}(A_2|B_2)$.
}
It is worth noting that these conditional entropies vanish when the corresponding conditional probabilities are $0$ or $1$. 
When both conditional entropies exist, the criteria $\mathcal{I}_{k}=\mathcal{I}_{A_{1}B_{1}}^{(k)}+\mathcal{I}_{A_{2}B_{2}}^{(k)}$ can be expressed as,
\begin{small}
\begin{align}
\mathcal{I}_{1}&=\frac{X(1-G_1)+X(1-G_2\sin(2\theta))-X(1-G_1\cos(2\theta))}{2\mathrm{\ln}(2)},\nonumber\\
\mathcal{I}_{2}&=\frac{X(\frac{1-F_2}{2})+X(1-\frac{(1+F_1)\sin(2\theta)}{2})-X(1-\frac{(1+F_2)\cos(2\theta)}{2})}{2\mathrm{\ln}(2)},
\end{align}
\end{small}
{\color{black}where the function $X(a)$ is defined as $X(a)=a\ln(a)+(2-a)\ln(2-a)$.}
We examine the sharing of quantum entanglement by checking whether both $\mathcal{I}_{1}$ and $\mathcal{I}_{2}$ exceed 1 simultaneously. It can be shown that, for any fixed pair of parameters $(G_1,G_2)$, the quantity $\min\{\mathcal{I}_{1},\mathcal{I}_{2}\}$ increases monotonically with $\theta$. Consequently, the maximal violation occurs at $\theta=\frac{\pi}{4}$. 
By further optimizing over $(G_1,G_2)$, we find that the largest attainable value $\max\{\min\{\mathcal{I}_{1},\mathcal{I}_{2}\}\}$ is $1.089$
achieved at $G_1=0.994$ and $G_2=0.397$.

\begin{figure*}[htbp]
	\centering 
	\subfloat[Mutual information]{\label{01-i}
		{\includegraphics[width=0.97\textwidth]{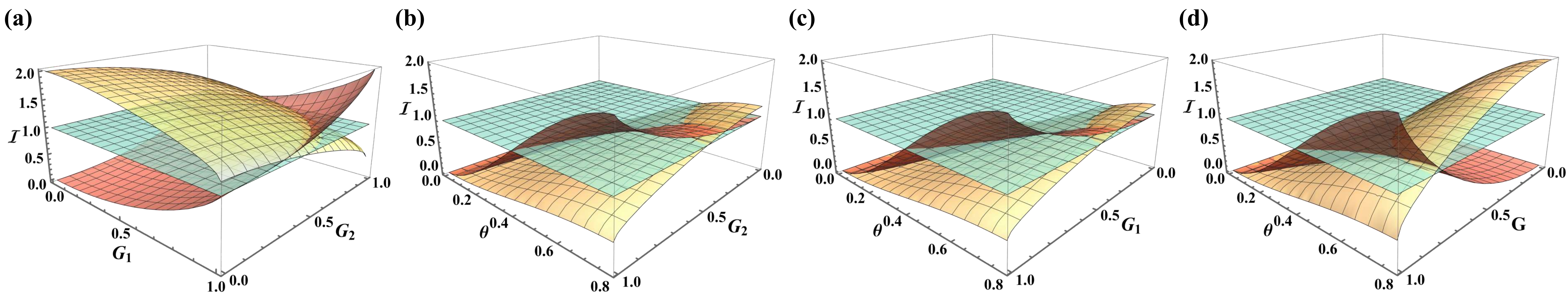}}} 
	
	\subfloat[Sum of conditional probabilities]{\label{01-s}
		{\includegraphics[width=0.97\textwidth]{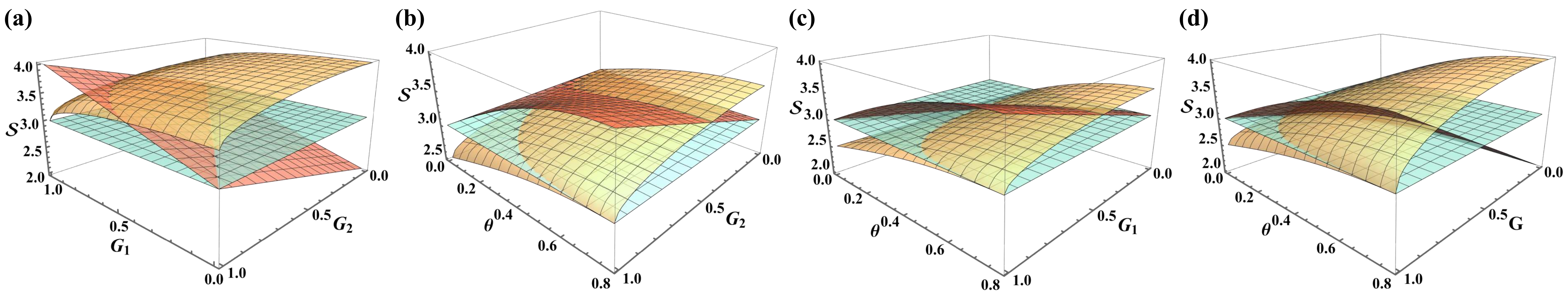}}} 
	
	\subfloat[Pearson correlation]{\label{01-c}
		{\includegraphics[width=0.97\textwidth]{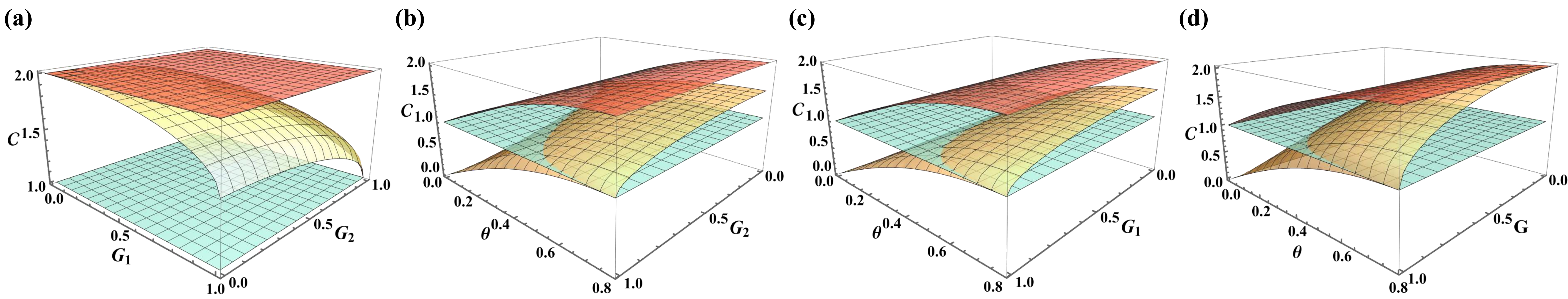}}} 
	\caption{\small{Unilateral entanglement sharing via weak measurement strategy:   
   {\color{black}
   (a) Case~1: the initial state is the maximal entangled state.
   (b) Case~2: the initial state is the partial entangled state with $ G_1 = 1 $.
   (c) Case~3: the initial state is the partial entangled state with \( G_2 = 1 \).
   (d) Case~4: the initial state is the partial entangled state with symmetric sharpness parameters \( G_i=G\), $i\in\{1,2\}$.}
    }}
	\label{oneside-weak}
\end{figure*}

For the maximally entangled state ($\theta=\frac{\pi}{4}$), the classical bound $\mathcal{I}_{1}=1$ corresponds to
$X(1-G_1)+X(1-G_2)=2\ln(2) $, which indicates violation ($\mathcal{I}_{1}>1$) throughout $0 <G_{1,2}< 1$.  
Similarly, for $\mathcal{I}_{2}$ the condition $\mathcal{I}_{2}=1$ is given by $X(\tfrac{1-F_1}{2})+X(\tfrac{1-F_2}{2})=2\ln(2) $,
with violation ($\mathcal{I}_{2}>1$) confined within this curve, bounded by the critical points $\{G_1,G_2\}=\{1,0.46\}$ and $\{0.46,1\}$.  
Consequently, simultaneous violation ($\mathcal{I}_{1},\mathcal{I}_{2}>1$) arises in the overlap of these two constraints, as shown in Fig.~\ref{01-i}(a). The optimal trade-off satisfies
$X(1-G_1)+X(1-G_2)=X(\tfrac{1-F_2}{2})+
X(\tfrac{1-F_1}{2})$.

Similarly, we investigate an extreme asymmetric measurement strategy. Supposed that one of the measurements is a strong measurement, which means that its information gain coefficient is 1.
In this case, the maximum value of $\max\{\min\{I_1,I_2\}\}$ attains $1.081$ when the information gain paprameter of the other measurement equals $0.332$. Specifically, for $G_1=1$, both inequalities $I_1>1$ and $I_2>1$ are simultaneously satisfied within the region $\theta \in (0.628,\frac{\pi}{4}]$ and $G_2 \in (0,0.467)$, as illustrated in Fig.~\ref{01-i}(b). Conversely, for $G_2=1$, both inequalities hold in the region $\theta \in (0.679,\frac{\pi}{4}]$ and $G_1 \in (0,0.467)$, 
as shown in Fig.~\ref{01-i}(c).  For the symmetric strategy ($G_1=G_2$), the maximum of $\max\{\min\{I_1,I_2\}\}$ is $1.06$, obtained at {\color{black}$G_{1,2}=G=0.8$}, with a narrower violation region, as illustrated in Fig.~\ref{01-i}(d).

We investigate entanglement sharing under the criteria of the sum of conditional probabilities. By substituting the joint probabilities into Eq.~(\ref{sab}), we obtain
\begin{align}\label{sum of probability}
	\mathcal{S}_1 &= 2 + G_1 + G_2 \sin(2\theta), \nonumber \\
       \mathcal{S}_2 &=2+ \tfrac{1}{2}(t(F_2) +t(F_1)\sin(2\theta)),
\end{align}
where $t(F_i)=1 + F_i$. The quantity $\max\{\min(\mathcal{S}_1,\mathcal{S}_2)\}$ always reaches its peak at $\theta=\frac{\pi}{4}$, 
with a maximum value of $\frac{18}{5}$ at {\color{black}$G_{1,2}=G=\frac{4}{5}$} [Fig.~\ref{01-s}(d)]. 
For maximally entangled states, when $G_1>1-G_2$, both criteria exceed $3$, as shown in Fig.~\ref{01-s}(a). With one sharpness parameter fixed at unity and the other reduced to $\frac{1}{\sqrt{5}}$, the asymmetric strategy still allows the maximally entangled state to reach $\max\{\min(\mathcal{S}_1,\mathcal{S}_2)\}=3+\frac{1}{\sqrt{5}}$. Furthermore, we analyze the constraints for partially entangled states to exhibit double violations under the asymmetric strategy. 
When $G_1=1$, both criteria exceed $3$ if 
$G_2 < \sqrt{2\sin(2\theta)-\sin^2(2\theta)}$, as illustrated in Fig.~\ref{01-s}(b). 
Conversely, when $G_2=1$, if 
$1-\sin(2\theta)< G_1 < \sqrt{1 - \left(\tfrac{1}{\sin(2\theta)}-1\right)^2}$, 
both criteria again exceed $3$, as shown in Fig.~\ref{01-s}(c). 
Clearly, the constraints in these two cases are asymmetric.

{\color{black}
We now turn to the Pearson correlation coefficient, which can be equivalently expressed in terms of joint probabilities, yielding
}
\begin{align}\label{Pearson correlation}
	\mathcal{C}_1 =& 1 + \sin(2\theta), \nonumber \\
        \mathcal{C}_2 =&\frac{t(F_1)\sin(2\theta)}{2} +\frac{t(F_2)\sin(2\theta)}{\sqrt{f(F_2)}},
\end{align}
where $f(F_i)=4 -t^2(F_i)\cos^2(2 \theta)$.
It can be observed that at $\theta = \frac{\pi}{4}$, the quantity $\max\{\min\{\mathcal{C}_{1}, \mathcal{C}_{2}\}\}$ approaches 2 as both information gain parameters tend to zero, though $\mathcal{C}_1$ becomes singular exactly at zero sharpness due to a vanishing denominator.
For a maximally entangled shared state, $\max\{\min\{\mathcal{C}_{1},\mathcal{C}_{2}\}\}$ always exceed 1, thereby enabling a complete verification of entanglement sharing, as depicted in Fig.~\ref{01-c}(a).
Next, we consider the scenario with an asymmetric measurement strategy. For \( G_1 = 1 \), the quantity $\max\{\min\{\mathcal{C}_{1},\mathcal{C}_{2}\}\}$ can exceed 1 (thus witnessing entanglement sharing) when 
\( G_2 < \sqrt{1 - \left(\sqrt{\frac{4}{(\frac{2\sin(2\theta)}{2 - \sin(2\theta)})^2 + \cos(2\theta)^2}} - 1\right)^2} \),
where \( \theta\in(0,\frac{\pi}{4}] \), 
as depicted in Fig.~\ref{01-c}(b).
Similarly, for \( G_2 = 1 \), the same quantity exceeds 1 if 
\( G_1 < \sqrt{1 - \left(\frac{2}{\sin(2\theta)} - \frac{2}{\sqrt{3 + \sin(2\theta)^2}} - 1\right)^2} \) 
with \( 0.353 < \theta \leq \frac{\pi}{4} \), as depicted in Fig.~\ref{01-c}(c).
{\color{black}
We find that both the $\mathcal{S}_k$ and $\mathcal{C}_k$ criteria successfully certify entanglement sharing across the full parameter regime $0 < G_i < 1$ under asymmetric measurement strategy scenarios.
Among the criteria $\{\mathcal{I}_k,\mathcal{S}_k,\mathcal{C}_k\}$, the Pearson correlation criterion proves to be the most effective.
}



In the study of entanglement sharing, the post-measurement state after Alice$_1$ perform measurements, can be written as, $ \rho'=(\hat{A}_{1,m}^{a_1} \otimes \mathbb{I}) \, \rho \, (\hat{A}_{1,m}^{a_1} \otimes \mathbb{I})^\dagger $.
The mixedness of this state can be quantified using the standard definition, yielding 
$\mathrm{Tr}[\rho'^2] =\frac{1}{8}(2+F_1^2+F_2^2+t(F_1)t(F_2))\le1$.
Furthemore, by applying the PPT criterion---which provides a necessary and sufficient condition---we can test whether the post-measurement state remains entangled. The PPT analysis confirms that the state preserves entanglement both before and after the intermidate measurement, in agreement with the results obtained from the sum of conditional probabilities and the Pearson correlation coefficient. Further details are provided in Appendix~\ref{Appdendix-A}.

\subsection{Unilateral Sharing Via PPM Strategy}

\begin{figure*}[htbp]
	\centering 
	\subfloat[Mutual information]{\label{02-i}
		{\includegraphics[width=0.97\textwidth]{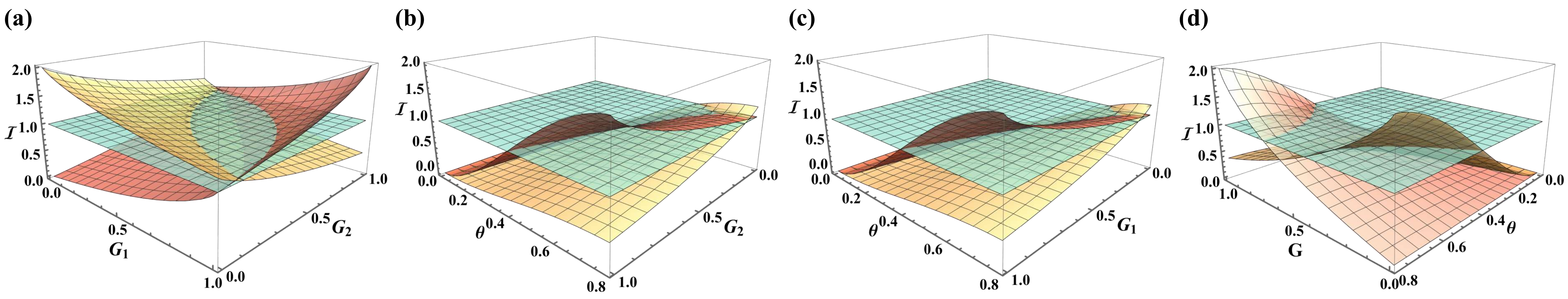}}} 
	
	\subfloat[Sum of conditional probabilities]{\label{02-s}
		{\includegraphics[width=0.97\textwidth]{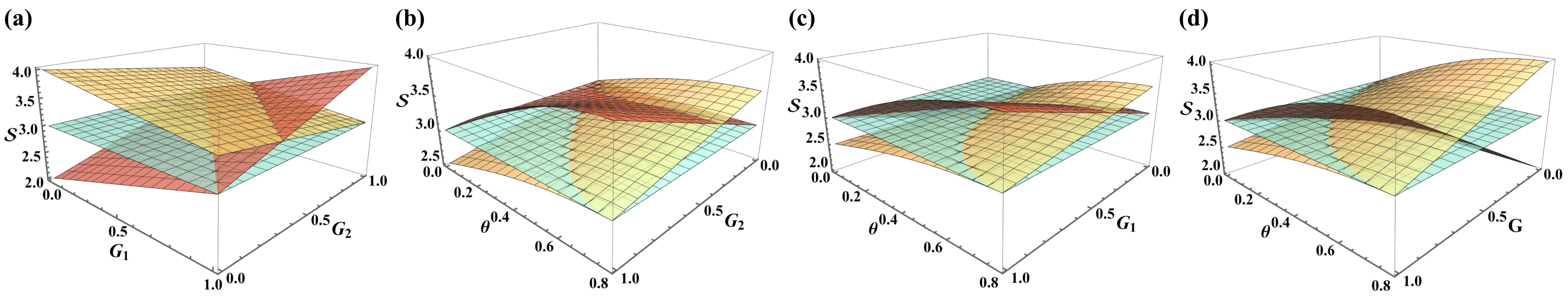}}} 
	
	\subfloat[Pearson correlation]{\label{02-c}
		{\includegraphics[width=0.97\textwidth]{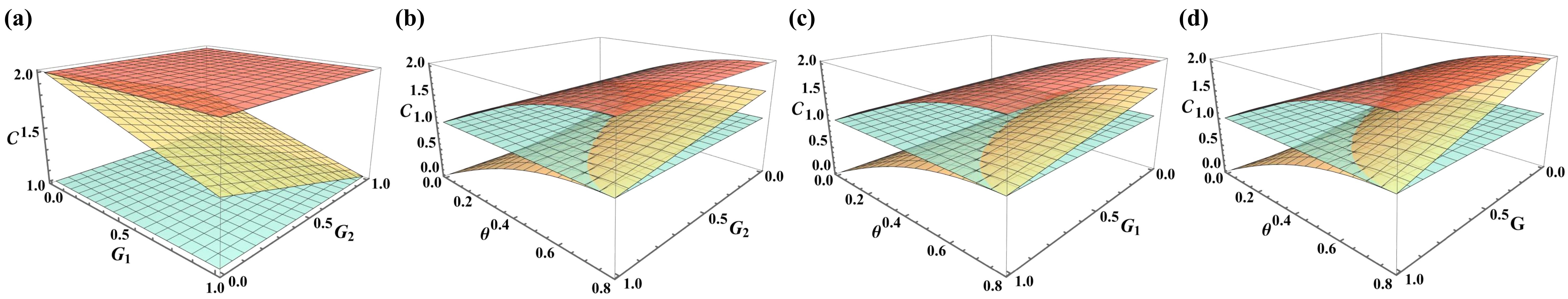}}} 
	\caption{\small{Unilateral entanglement sharing via PPM strategy: 
   (a) Case~1: the initial state is the maximal entangled state.
   (b) Case~2: the initial state is the partial entangled state with $ G_1 = 1 $.
   (c) Case~3: the initial state is the partial entangled state with \( G_2 = 1 \).
   (d) Case~4: the initial state is the partial entangled state with symmetric sharpness parameters \( G_i=G\), $i\in\{1,2\}$.}}
	\label{oneside-ppt}
\end{figure*}


We now analyze entanglement sharing via the PPM strategy, following a methodology analogous to the previous section.
Since the information gain parameter $G_i=\alpha_{\hat{A}_{1,i}}$ ($i \in \{1, 2\}$) in this case, and the degree to which the system remains undisturbed is represented by the factor $F_i =1 -\alpha_i= 1-G_i$, the mutual information criteria $\mathcal{I}_{k}$ ($k\in\{1,2\}$) can be written as,
\begin{small}
\begin{align}
	\mathcal{I}_{1}=&1-\sin(\theta)^2+ \frac{ X(G_2(1 - \sin(2\theta))) + X(G_2(1 + \sin(2\theta)))}{4 \mathrm{\ln}(2)}\nonumber\\
	&-\frac{X(G_1(1-\cos(2\theta)))+X(G_2)-X(2G_1)\sin(\theta)^2}{2\mathrm{\ln}(2)},\nonumber\nonumber\\
       \mathcal{I}_{2}=&\frac{X(\frac{1-F_2}{2})+X(1 -\frac{( 1+F_1) \sin(2\theta)}{2})-X(1 -\frac{(1+ F_2) \cos(2\theta)}{2})}{2 \mathrm{\ln}(2)}.
\end{align}
\end{small}
{\color{black}
Notably, the expression for $\mathcal{I}_2$ derived from the PPM strategy coincides with that from the weak measurement strategy, whereas the functional form of $\mathcal{I}_{1}$ differs between the two strategies.}
It can be observed that the quantity $\max\{\min\{\mathcal{I}_{1},\mathcal{I}_{2}\}\}$ is always achieved at $\theta = \frac{\pi}{4}$, with optimal parameters $G_1 = 0.125$ and $G_2 = 0.857$ yielding a global maximum of approximately 1.05.
For a maximally entangled shared state, the condition $\mathcal{I}_1 = 1$ corresponds to $X(2G_1)+X(2G_2)-2X(G_1)-2X(G_2)=2\ln(2)$
while the curve for $\mathcal{I}_2 = 1$ satisfies $X(\frac{1-F_1}{2}) + X(\frac{1-F_2}{2}) = 2 \mathrm{\ln}(2)$. 
The region where both $\mathcal{I}_1$ and $\mathcal{I}_2$ exceed 1, defined for $G_i\in(0, 0.116)$, is bounded by these curves, as illustrated in Fig.~\ref{02-i}(a).
For a partially entangled state under an asymmetric measurement strategy with one information gain parameter fixed at unity, the quantity $\max\{\min\{\mathcal{I}_1, \mathcal{I}_2\}\}$ reaches a maximum value of 1.043 when the other parameter is set to 0.083.
Moreover, it exceeds unity when this parameter lies within the interval $(0, 0.116)$.
For the case $G_1 = 1$, the inequality $\max\{\min\{\mathcal{I}_1, \mathcal{I}_2\}\} > 1$ is satisfied for $\theta \in (0.67, \tfrac{\pi}{4}]$, as shown in Fig.~\ref{02-i}(b).
Similarly, for $G_2 = 1$, the violation holds over $\theta \in (0.711, \tfrac{\pi}{4}]$, as depicted in Fig.~\ref{02-i}(c).
However, no simultaneous violation occurs under symmetric settings ({\color{black}$G_{1,2} = G$}), as illustrated in Fig.~\ref{02-i}(d).
It is noteworthy that the mutual information criterion using the weak measurement protocol outperforms the PPM strategy in this scenario.

{\color{black}
We analyze entanglement sharing in this scenario using the sum of conditional probabilities. The expression for $S_k$ under the PPM strategy coincides with that from weak measurements, as shown in Eq.~\eqref{sum of probability}. 
While, it is necessary to emphsized that 
{\color{black}$G_i$ and $F_i$ satisfy $F_i+G_i=1$, }
indicating the fundamental trade-off between information gain and state disturbance.}
Within this framework, when $G_1 = \frac{4}{3} - G_2$ and $\theta = \frac{\pi}{4}$, the value of $\max\{\min\{\mathcal{S}_1,\mathcal{S}_2\}\}$ attains its maximum value of $\frac{10}{3}$. 
For the maximally entangled state, the double violations of both criteria occur for $G_1 > 1 - G_2$, demonstrating entanglement sharing, as shown in Fig.~\ref{02-s}(a). Furhter the asymmetric measurement strategies are analyzed. For $G_1 = 1$, simultaneous violations occur when $G_2 < \sin(2\theta)$,  as depicted in Fig.~\ref{02-s}(b). 
Conversely, for $G_2 = 1$, double violations arise when  $1 - \sin(2\theta) < G_1 < 2 - \frac{1}{\sin(2\theta)}$  and $\frac{1 }{2}\arcsin\!\left(\frac{-1+\sqrt{5}}{2}\right) < \theta \leq \frac{\pi}{4}$, 
as shown in Fig.~\ref{02-s}(c).  In the symmetric case ({\color{black}$G_{1,2} = G$}), simultaneous violations occur when 
$\frac{1}{1+\sin(2\theta)} < G < \frac{2\sin(2\theta)}{1+\sin(2\theta)}$ 
with $\frac{\pi}{12} < \theta \leq \frac{\pi}{4}$, 
as illustrated in Fig.~\ref{02-s}(d).

Similarly, under the Pearson correlation coefficient criterion, 
the form of $C_k$ obtained from the PPM strategy coincides with that from weak measurements as is shown in Eq.~\eqref{Pearson correlation}. Nevertheless, the trade-off between information gain and state disturbance differs between the two strategies.  For $\theta = \frac{\pi}{4}$, we find that $\max\{\min\{\mathcal{C}_1,\mathcal{C}_2\}\}$ approaches $2$ as the information gain $G_i$ tends to zero. Thus, for the maximally entangled state, this value always exceeds unity, which exhibit entanglement sharing in the entire region, as shown in Fig.~\ref{02-c}(a).
Subsequently, we analyze the extreme asymmetric measurement strategy via PPM strategy.
For the case \( G_1 = 1 \), the two criteria simultaneously exceed 1—indicating entanglement sharing—when \( G_2 < 2 - \sqrt{\frac{4}{\left(\frac{2\sin(2\theta)}{\sin(2\theta) - 2}\right)^2 + \cos(2\theta)^2}} \) for \( \theta\in(0, \frac{\pi}{4} ]\), as shown in Fig.~\ref{02-c}(b).
Similarly, for \( G_2 = 1 \), both criteria simultaneously exceed 1 when \( G_1 < 2(1+\sqrt{\frac{1}{3 + \sin(2\theta)^2}} - \frac{1}{\sin(2\theta)}) \) for \(  \theta \in(0.353 ,\frac{\pi}{4}] \), as depicted in Fig.~\ref{02-c}(c).
{\color{black}
In the PPM strategy, both \(\mathcal{S}_k\) and \(\mathcal{C}_k\) certify entanglement sharing throughout the entire parameter range \( G_i \in (0,1) \), in agreement with the weak measurement scenario. Among the three criteria \(\{\mathcal{I}_k, \mathcal{S}_k, \mathcal{C}_k\}\), the Pearson correlation coefficient proves to be the most reliable indicator.

}
Furthermore, under this strategy Alice’s post-measurement state can be explicitly represented and analyzed, as detailed in Appendix~\ref{Appdendix-A}. 
The mixedness of the state is given by $\frac{1}{16}[8+F_1t(F_1)+(t(F_1)+2t(F_2))F_2+(1-F_1)(t(F_1)+t(F_2))\cos(4\theta)]<1$.  
Applying the PPT criterion, we find that the minimum eigenvalue of the partially transposed density matrix $\rho^T$ is always negative, thereby confirming that Alice$_1$’s post-measurement state remains entangled for all parameter values. 
This result further corroborates the conclusions obtained from both the sum of conditional probabilities and the Pearson correlation coefficient criteria. 
Additional details are provided in Appendix~\ref{Appdendix-A}.



\section{Bilateral Sequential Entanglement Sharing}\label{twoside}


After analyzing the unilateral scenario, we turn to the bilateral sequential scenario, where both particles undergo sequential measurements (see Fig.~\ref{general}). 
Here the entanglement between Alice$_1$--Bob$_1$ and Alice$_2$--Bob$_2$ is quantified through three criteria, $\mathcal{I}_k$, $\mathcal{S}_k$, and $\mathcal{C}_k$ ($k=\{1,2\}$), which follow directly from the joint probability distribution in Eq.~\eqref{joint measurement result-1}.\\

\subsection{Bilateral Sharing via Weak Measurement Strategy}


\begin{figure*}[htbp]
\centering
\includegraphics[width=0.8\textwidth]{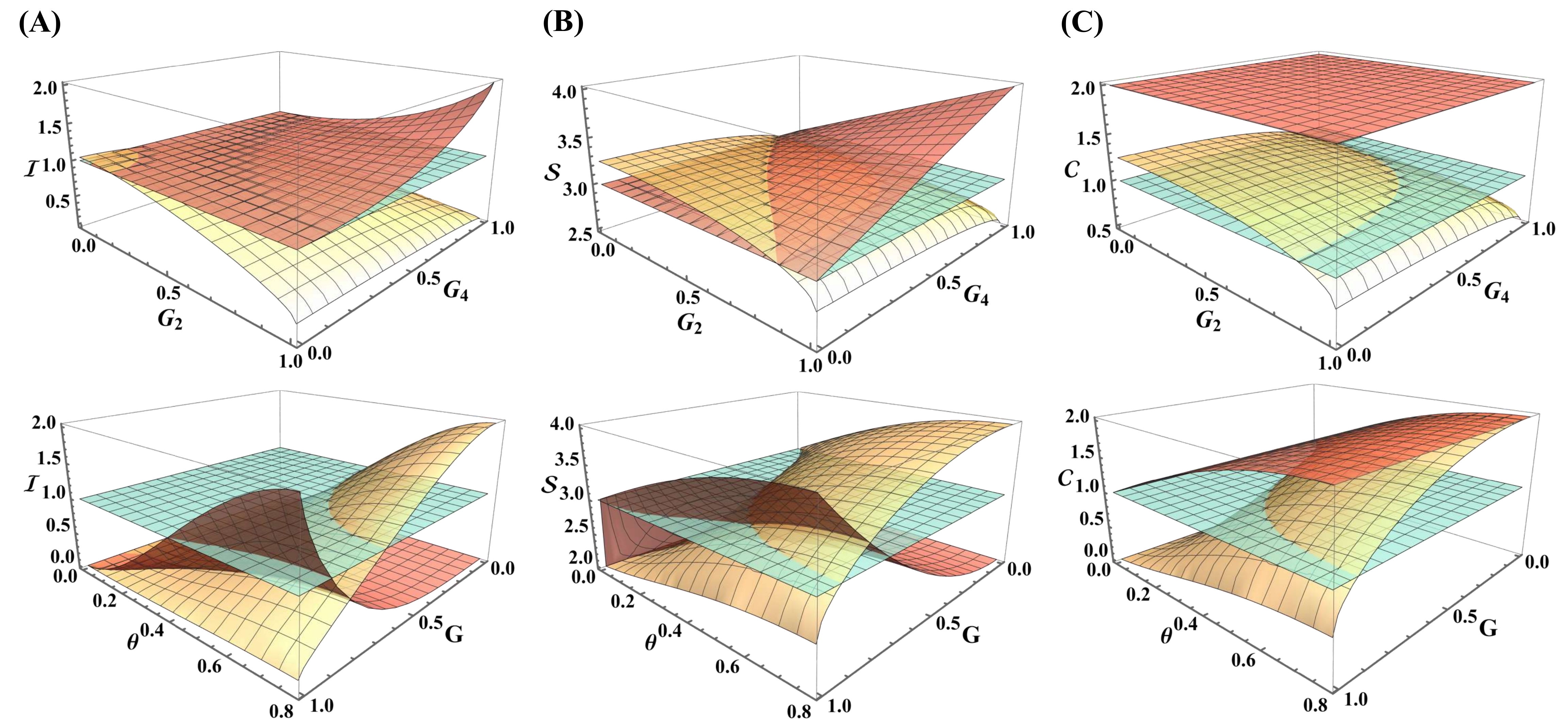}
\caption{
 Bilateral entanglement sharing via weak measurement strategy:  Subfigures (A), (B), and (C) correspond to the criteria $\{\mathcal{I}_{k} , \mathcal{S}_{k}, \mathcal{C}_{k}\}$, respectively. Top panel: The case where $G_1=1$, $G_3=1$. Bottom panel: All parameters equal.
}\label{03-csi}
\end{figure*}

We first analyze entanglement sharing in the bilateral sequential scenario based on a weak measurement strategy. 
Without loss of generality, the information-gain parameters defined as 
$G_i=\{\eta_{\hat{A}_{1,1}}, \eta_{\hat{A}_{1,2}}, \eta_{\hat{B}_{1,1}}, \eta_{\hat{B}_{1,2}}\}$ 
with $i\in\{1,2,3,4\}$, and the corresponding perturbation factors 
$F_i=\sqrt{1-G_i^2}$. 
In this case, the mutual-information criterion $\mathcal{I}_k$ can be written as,
\begin{widetext}
\begin{align}
\mathcal{I}_1=&\frac{1}{4 \mathrm{\ln}(2)}(2X(1-G_2G_4\sin[2\theta])+X(1-G_1G_3+(-G_1+G_3)\cos(2\theta),2(1-G_1G_3))-2X(1+G_3\cos(2\theta))\nonumber\\
&+X(1+G_1G_3+(G_1+G_3)\cos(2\theta),2(1+G_1G_3))-2X(1-G_1\cos(2\theta)))\nonumber\\
\mathcal{I}_2=&-2+\frac{1}{2\mathrm{\ln}(2)}(X(1 + \frac{(1 + F_1)(1 +F_3) \sin(2 \theta)}{4})-X(1-\frac{(1+F_2)\cos(2\theta)}{2})-X(1 + \frac{(1 + F_4) \cos(2 \theta)}{2})\nonumber\\
&+\frac{1}{8}(X(4-(F_2+1)(F_4+1) - 2 (F_2 - F_4) \cos(2 \theta),8-4(1+F_2)\cos(2\theta))\nonumber\\
&+X(4-(F_2+1)(F_4+1) + 2 (F_2 - F_4) \cos(2 \theta),8+4(1+F_2)\cos(2\theta)))).
\end{align}
\end{widetext}
where $X(a,b)=a\mathrm{\ln}[a]+(b-a)\mathrm{\ln}[b-a]$. 
Assuming a maximally entangled initial state and fixing \( G_1 = G_3 = 1 \), we find that 
\( \mathcal{I}_{1} > 1 \) whenever \( G_2, G_4 \in (0,0.14) \), while \( \mathcal{I}_{2} > 1 \) holds in this regime as well. 
Nevertheless, no simultaneous violation occurs in the following cases: 
(i) \( \theta = \frac{\pi}{4} \), \( G_1 = G_3 \), \( G_2 = G_4 \); 
(ii) \( \theta = \frac{\pi}{4} \), \( G_1 = G_4 \), \( G_2 = G_3 \); and 
(iii) {\color{black} \( G_{1,2,3,4} = G \). }
As an illustration, in case (iii) with \( G= \tfrac{4}{5} \) and \( \theta = \tfrac{\pi}{4} \), 
the maximum value of \( \max\{\min\{\mathcal{I}_1,\mathcal{I}_2\}\} \) reaches only 0.64, 
well below unity, indicating that entanglement sharing cannot be certified, 
as shown in Fig.~\ref{03-csi}(A).

We investigate entanglement sharing under the criteria of
the sum of conditional probabilities. By substituting the joint
probabilities into Eq.~\eqref{sab}, we obtain
\begin{align}
\mathcal{S}_{1}=& 2+G_2G_4\sin(2\theta)+\frac{G_1G_3\sin^2(2\theta)}{1-G_3^2\cos^2(2\theta)},
\nonumber\\
    \mathcal{S}_{2} =&2+ \frac{1}{4}t(F_1)t(F_3)\sin(2 \theta)
        + \frac{t(F_2)t(F_4)\sin^2(2 \theta)}{f(F_4)}.
\end{align}
As shown in Fig.~\ref{03-csi}(B),
we find that the maximum value of $\max\{\min\{\mathcal{S}_1,\mathcal{S}_2\}\}$ reaches $\frac{82}{25}(\approx 3.28)$ when \( G_{1,2,3,4} =\frac{4}{5} \) and \( \theta=\frac{\pi}{4} \). 
{\color{black}
Under the symmetric condition \(G_i = G\) ($F=\sqrt{1-G^2}$), the curves corresponding to $\mathcal{S}_1=3$ and $\mathcal{S}_2=3$ are given by,
$G^2\sin(2 \theta) =\frac{1- G^2}{1- G^2 \cos^2(2 \theta)} $ 
and
$\frac{1}{4}t^2(F)\sin(2\theta)+\frac{t^2(F)\sin^2(2\theta)}{f(F)}= 1$.
Both of them simultaneously exceed 3 within the parameter range \( G\in \left(\frac{1}{\sqrt{2}},\sqrt{2(\sqrt{2} - 1)}\right) \). 
}
For the specific asymmetric case with \( G_1 = G_2 = 1 \) and \( \theta = \frac{\pi}{4} \), the maximum value \(\max\{\min\{\mathcal{S}_1,\mathcal{S}_2\}\}\) reaches $\frac{2(34+\sqrt{31})}{25}(\approx 3.17)$ at \( G_2 = G_4 = \sqrt{\frac{2\sqrt{31}-7}{25}} \). Here, \(\mathcal{S}_1 = 3 + G_2 G_4 >3\), while \(\mathcal{S}_2=1 \) implies $t(F_2)t(F_4)=3$.
Entanglement can be observed when 
$t(F_2)t(F_4)>3$.

We turn to the Pearson correlation coefficient, which
can be equivalently expressed in terms of joint probabilities,
yielding
\begin{align}
	\mathcal{C}_1 =& 1 + \sin(2\theta), \nonumber \\
	\mathcal{C}_2 =& \frac{1}{4}t(F_1)t(F_3) \sin(2 \theta)
	+ \frac{t(F_2)t(F_4) \sin^2(2 \theta)}{\sqrt{f(F_2) f(F_4)}}.
\end{align}
Here, \( \mathcal{C}_1\) consistently exceeds 1. 
We focus on the maximally entangled state with \( \theta = \tfrac{\pi}{4} \). 
In the limit where all information–gain parameters \( \{G_i\} \) tend to zero, 
\(\max\{\min\{\mathcal{C}_1,\mathcal{C}_2\}\}\) approaches its maximum value of 2. 
For symmetric scenario with \( G_i = G \), the condition \(\mathcal{C}_2=1\) is satisfied by 
{\color{black}$\frac{1}{4} t^2(F)\sin(2\theta) + \frac{t^2(F)\sin^2(2\theta)}{f(F)} = 1$,}
and \(\max\{\min\{\mathcal{C}_1,\mathcal{C}_2\}\} > 1\) is achieved for 
\( G \in \bigl(0, \sqrt{2(\sqrt{2}-1)}\bigr) \). 
We further consider an asymmetric scenario with \( G_1 = G_3 = 1 \). 
In this case, \(\mathcal{C}_1\) remains fixed at 2, while the condition \(\mathcal{C}_2=1\) requires \( t(F_2)t(F_4) =3\). 
Entanglement sharing is thus certified whenever \( t(F_2)t(F_4) > 3 \). 
The results for both symmetric and asymmetric scenarios are summarized in Fig.~\ref{03-csi}(C). Notably, although the $\mathcal{S}_k$ and $\mathcal{C}_k$ criteria are derived independently, they yield identical parameter ranges for entanglement sharing under asymmetric measurement conditions.

In the setting of bilateral entanglement sharing with weak measurements, the Pearson correlation coefficient proves to be the most sensitive indicator.  We also examine the mixedness of the post-measurement states, and apply the PPT criterion to assess their entanglement. For the symmetric choice of parameters \(\{ G_1 = G_3 = 1, \, \theta = \tfrac{\pi}{4} \}\), the resulting states may become separable after measurement. These results are also consistent with the entanglement detection established by the three criteria discussed above. Further technical details are provided in Appendix~\ref{Appdendix-A.2}.

\subsection{Bilateral Sharing Via PPM Strategy}


\begin{figure*}[htbp]
	\centering
	\includegraphics[width=0.8\textwidth]{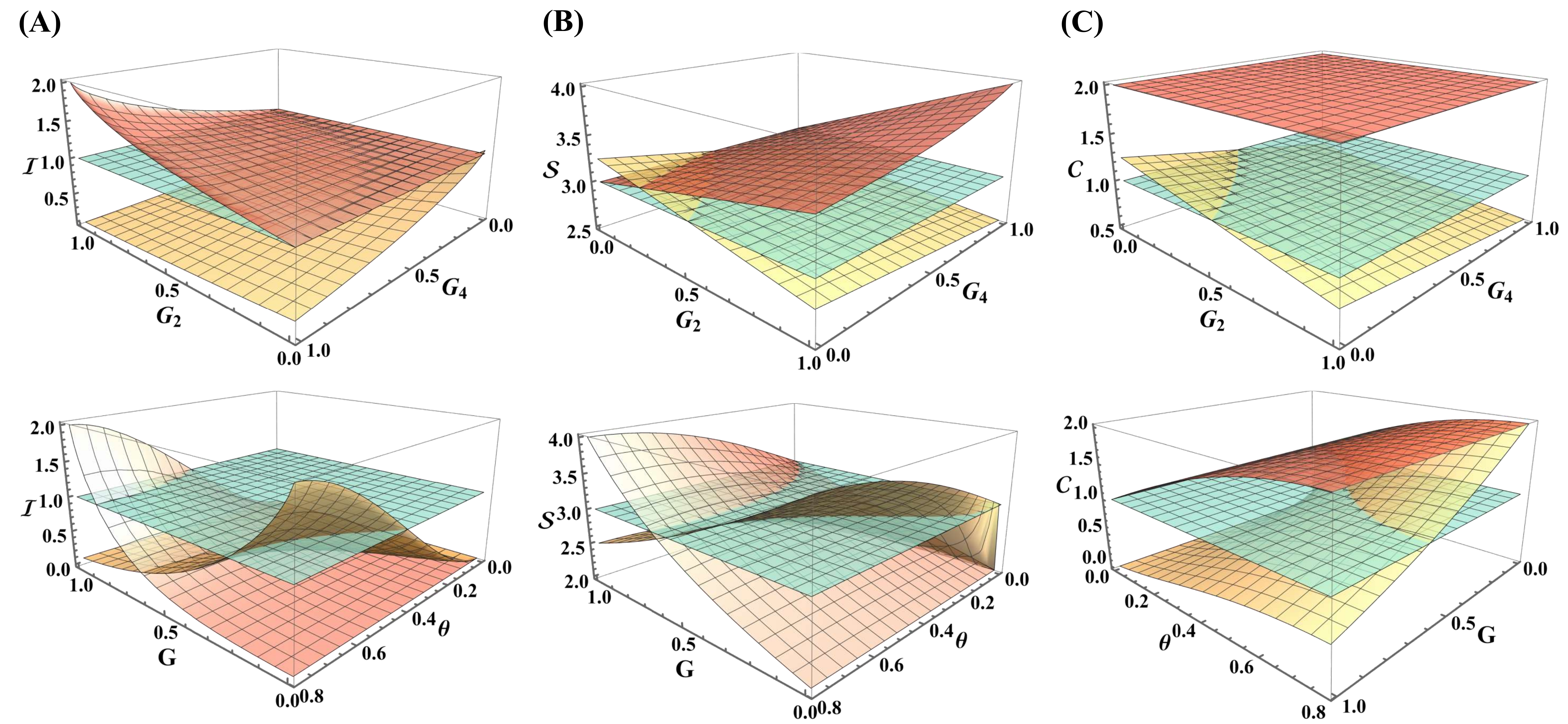}
	\caption{
			Bilateral entanglement sharing via PPM strategy:  Subfigures (A), (B), and (C) correspond to the criteria $\{\mathcal{I}_{k} , \mathcal{S}_{k}, \mathcal{C}_{k}\}$, respectively. Top panel: The case where $G_1=1$, $G_3=1$. Bottom panel: All parameters equal.
	}\label{04-csi}
\end{figure*}

We explore the entanglement sharing based on the PPM strategy in the bilateral sequential scenario. To unify the symbols, the information gain parameters are defined as $\{\alpha_{\hat{A}_{1,1}},\alpha_{\hat{A}_{1,2}},\alpha_{\hat{B}_{1,1}},\alpha_{\hat{B}_{1,2}}\}$, which we denote as $\{G_i\}$ for $i = \{1, 2, 3, 4\}$, and the corresponding perturbation factors $F_i = 1 - G_i$.
In this case, the mutual-information criterion $\mathcal{I}_k$ ($k\in\{1,2\}$) can be given as,
\begin{widetext}	
\begin{small}
	\begin{align}
		\mathcal{I}_1=&1+\frac{1}{4\mathrm{\ln}(2)}(G_4 X(G_2 (\cos\theta + \sin^2\theta)))+X(G_2 \left( 2 - G_4 - G_4 \sin2 \theta \right),4-2G_4)-2(2-G_4)\mathrm{\ln}[2-G_4]-2X(G_2)\nonumber\\
		&-2X(G_1(1-\cos2\theta)))
		+\frac{1}{\mathrm{\ln}(2)} (G_3 \sin^2\theta X(a,1) +  X(G_1 (1 - G_3) \sin^2\theta,1-G_3\sin^2\theta )  - (1-G_3\sin^2\theta ) \mathrm{\ln} [1-G_3\sin^2\theta ])\nonumber\\
        \mathcal{I}_2=&-2+\frac{1}{2\mathrm{\ln}(2)}(X(1 + \frac{(1 + F_1)(1 +F_3) \sin2 \theta}{4})-X(1-\frac{(1+F_2)\cos2\theta}{2})-X(1 + \frac{(1 + F_4) \cos2 \theta}{2})\nonumber\\
        &+\frac{1}{8}(X(4-(F_2+1)(F_4+1) - 2 (F_2 - F_4) \cos2 \theta,8-4(1+F_2)\cos2\theta)\nonumber\\
        &+X(4-(F_2+1)(F_4+1) + 2 (F_2 - F_4) \cos2 \theta,8+4(1+F_2)\cos2\theta))).
	\end{align}
\end{small}
\end{widetext}
{\color{black}
We observe that the expression for $\mathcal{I}_2$ derived from the PPM strategy coincides with that from the weak measurement strategy, while the form of $\mathcal{I}_{1}$ differs between the two strategies.}
Assuming a maximally entangled initial state and fixing \( G_1 = G_3 = 1 \), we find that 
\( \mathcal{I}_{1} > 1 \) for \( G_2, G_4 \in (0,0.02) \), while \( \mathcal{I}_{2} > 1 \) also holds in this regime. 
Compared with the weak measurement strategy, the region of double violation is substantially narrower and the violation strength significantly reduced. 
Moreover, no simultaneous violation is observed under the following conditions: 
(i) \( \theta = \tfrac{\pi}{4}, \, G_1 = G_3, \, G_2 = G_4 \); 
(ii) \( \theta = \tfrac{\pi}{4}, \, G_1 = G_4, \, G_2 = G_3 \); and 
(iii) {\color{black}\( G_{1,2,3,4} = G \). }
As a representative example, in case (iii) with {\color{black}\( G= \tfrac{4}{5} \)} and \( \theta = \tfrac{\pi}{4} \), 
the maximum value of \( \max\{\min\{\mathcal{I}_1,\mathcal{I}_2\}\} \) is only 0.32, 
well below 1, indicating that entanglement sharing cannot be verified, as shown in Fig.~\ref{04-csi}(A).

We investigate entanglement sharing under the criteria of
the sum of conditional probabilities. By substituting the joint
probabilities into Eq.~(\ref{sab}), it gives
\begin{align}
     \mathcal{S}_1 =&2+\frac{G_2\sin(2\theta)}{2-G_4}+\frac{G_1\cos^2\theta}{1-G_3\sin^2\theta},
     \nonumber\\
       \mathcal{S}_{2} =&2+ \frac{1}{4}t(F_1)t(F_3)\sin(2 \theta)
        + \frac{t(F_2)t(F_4)\sin^2(2 \theta)}{f(F_4)}.
\end{align}
Notably, the expression for $\mathcal{S}_2$ derived from the PPM strategy coincides with that from the weak measurement strategy, whereas the form of $\mathcal{S}_{1}$ differs between the two strategies. 
{\color{black}We first analyze a symmetric measurement strategy ($G_i=G$).}
The maximum value of $\max\{\min\{\mathcal{S}_1,\mathcal{S}_2\}\}$ reaches \(2.937\), when {\color{black}\(G = 0.627\)} and \(\theta = 0.729\). Hence, entanglement sharing can not exist. We then turn to an asymmetric strategy with \(G_1 = G_3 = 1\) and \(\theta = \tfrac{\pi}{4}\). In this case, \(\max\{\min\{\mathcal{S}_1,\mathcal{S}_2\}\}\) attains a larger value of \(3.125\). Specifically, \(\mathcal{S}_1\) simplifies to  $\mathcal{S}_1 = 3 + \frac{G_2}{2 - G_4}$, which always exceeds 3. By contrast, \(\mathcal{S}_2 > 3\) holds when $G_2 < \frac{1 - 2G_4}{2 - G_4}$ {\color{black}($t(F_2)t(F_4)>3$)}, with the boundary conditions given by the intersections \(\{G_2,G_4\} = \{\tfrac{1}{2},0\}\) and \(\{0,\tfrac{1}{2}\}\). Thus, both inequalities are satisfied for \(G_2, G_4 \in (0, \tfrac{1}{2})\) under the constraint above. The corresponding parameter region is illustrated in Fig.~\ref{04-csi}(B).

{\color{black}
We analyze the Pearson correlation coefficient in the bilateral sequential scenario. Remarkably, the expression for \(\mathcal{C}_k\) obtained under the PPM strategy, where \(G_i + F_i = 1\), coincides with that derived from the weak measurement strategy, which satisfies \(G_i^2 + F_i^2 = 1\).
Notably, despite their distinct parameter constraints, one key conclusion remains consistent across both strategies: $\mathcal{C}_1$ consistently exceeds 1, and the quantity $\max\{\min\{\mathcal{C}_1,\mathcal{C}_2\}\}$ approaches its maximum value as all information gain parameters approach zero for a maximally entangled initial state.}
{\color{black}
Under the symmetric strategy with equal information gain parameters ($G_i=G$), when $\mathcal{C}_2=1$ yields $\frac{1}{4}t^2(F) \sin(2 \theta)(1 + \frac{4 \sin(2 \theta)}{f(F)}) = 1$.
The region where $\max\{\min\{\mathcal{C}_1,\mathcal{C}_2\}\}>1$ corresponds to \( G \in (0, 2 - \sqrt{2}) \) and \( \theta \in (0, \frac{\pi}{4}) \).
}
For the asymmetric case with \( G_1 = G_3 = 1 \) and \( \theta = \frac{\pi}{4} \), the quantity $\max\{\min\{\mathcal{C}_1,\mathcal{C}_2\}\}>1$ holds when \( G_2 < \frac{1 - 2G_4}{2 - G_4} \), within the domain \( G_2,G_4 \in \left(0, \frac{1}{2}\right) \). 
Remarkably, under asymmetric measurement conditions, the $\mathcal{S}_k$ and $\mathcal{C}_k$ criteria yield identical parameter ranges for entanglement sharing.
The results are presented in Fig.~\ref{04-csi}(C).

Similarly, under this strategy, the post-measurement state were explicitly  analyzed. For comparison, we examine entanglement using the PPT criterion under two specific parameter configurations: (i) all parameters equal, and (ii) $\theta = \frac{\pi}{4}$ with $G_1 = G_2 = 1$. The results indicate that the quantum states remain entangled within specific parameter regions. Both the sum of conditional probabilities and the Pearson correlation coefficient criteria identify entangled regions that fall within the range predicted by PPT. Further details are provided in Appendix~\ref{Appdendix-A.2}.

\section{Conclusion}\label{Con}

{\color{black}

For a two-qubit quantum system, we systematically investigate the classical correlations exhibited by complementary measurement results from different observer combinations in a sequential measurement scenario. We carefully evaluate three complementary correlation metrics: mutual information, sum of conditional probabilities, and Pearson correlation coefficient, for both one-sided and two-sided sequential measurement scenarios, employing both weak measurement and probabilistic projective measurement (PPM) strategies. We further demonstrate how these measures capture correlations from distinct observer pairs and show that when correlations in multiple pairs simultaneously exceed their respective thresholds, the presence of entanglement sharing is unambiguously certified.

We find that weak measurement strategies are more favorable than PPM for exhibiting entanglement sharing, regardless of the scenarios. In the unilateral sequential scenario, the value of $\max\{\min\{\mathcal{I}_1,\mathcal{I}_2\}\}$ reaches 1.089 under the weak measurement strategy, exceeding the 1.05 under the PPM strategy. For both strategies, the corresponding maximal double violation is achieved under the asymmetric sharpness parameter. For the sum of conditional probabilities, the value of $\max\{\min\{\mathcal{S}_1, \mathcal{S}_2\}\}$ reaches $ \frac{18}{5}$, surpassing the result of $\frac{10}{3}$ obtained with the PPM strategy. Notably, the strongest violation arises under symmetric configurations of the sharpness parameters. The Pearson correlation criterion can more comprehensively capture the entanglement sharing phenomenon exhibited in all measurement strategies and reaches its maximum value in the symmetric setting.
In bilateral scenarios, the mutual information criterion fails to characterize entanglement sharing. By contrast, both the sum of conditional probabilities and the Pearson correlation remain effective under appropriate measurement conditions. Remarkably, these two criteria reach their optimal performance under weak measurements with symmetric sharpness parameters. In the PPM strategy, however, the sum of conditional probabilities criterion requires an asymmetric configuration of sharpness parameters to achieve the maximal double violation, whereas the Pearson correlation attains its maximum only under symmetric configurations.

Overall, the Pearson correlation criterion more readily demonstrates entanglement sharing and exhibits strong robustness in both unilateral and bilateral sequence scenarios. The principles revealed in this study, particularly the trade-off between measurement disturbance and complementary correlation recovery, are expected to generalize to other quantum resource reuse problems beyond entanglement sharing. These findings lay the foundation for designing efficient quantum resource recovery protocols.
}



\section{Acknowledgment}

C.R. was supported by the National Natural Science Foundation of China (Grant No. 12575016, 12421005,12247105,12075245), Hunan provincial major sci-tech program (No. 2023ZJ1010), the Natural Science Foundation of Hunan Province (2021JJ10033), the Foundation Xiangjiang Laboratory (XJ2302001) and Xiaoxiang Scholars Program of Hunan Normal University. 


\appendix
\section{Entanglement witness by positive partial transpose (PPT) criterion}
\subsection{Unilateral scenario}\label{Appdendix-A}

{\color{black}

In the unilateral sequential scenario using weak measurements, the partial transpose of Alice$_1$'s post-measurement density matrix has four eigenvalues,
\begin{small}
\begin{align}
		e_1 &= \frac{1}{8} ( 3 + F_2) ( 1 - \sqrt{1 - \frac{8t(F_2)  \sin^2(2 \theta)}{\left( 3 +F_2  \right)^2}} ),\nonumber\\
		e_2 &= \frac{1}{8} ( 3 +F_2) ( 1 + \sqrt{1 - \frac{8t(F_2) \sin^2(2 \theta)}{( 3 +F_2  )^2}} ),\nonumber\\
		e_3 &= \frac{1}{8} ( 1 - F_2) ( 1 - \sqrt{1 + \frac{4 ( -G_1^2 +t(F_1)t(F_2) ) \sin^2(2 \theta)}{\left( 1 -F_2 \right)^2}} ),\nonumber\\
		e_4 &= \frac{1}{8} ( 1 -F_2 ) ( 1 + \sqrt{1 + \frac{4 \left( -G_1^2 + t(F_1)t(F_2) \right) \sin^2(2 \theta)}{( 1 - F_2)^2}} ),
\end{align}
\end{small}
which serve as a direct signature of entanglement in the system.
Since $e_1$, $e_2$, and $e_4$ are positive while $e_3$ is negative, the post-measurement state is always entangled.


For the PPM strategy, the four eigenvalues of the partial transpose of Alice$_1$'s post-measurement density matrix are
\begin{small}
\begin{align}
	&e_1 = \frac{1}{8} (4 - G_2) (1 - \sqrt{1 -\frac{8 (2 - G_2) \sin^2(2 \theta)}{(4 - G_2)^2}});\nonumber\\
	&e_2 = \frac{1}{8} (4 - G_2) (1 + \sqrt{1 - \frac{8 (2 - G_2) \sin^2(2 \theta)}{(4 - G_2)^2}});\nonumber\\
	&e_3 = \frac{1}{8} G_2 (1 - \sqrt{1 + \frac{4 (2 - G_1) (2 - G_1 - G_2) \sin^2(2 \theta)}{G_2^2}});\nonumber\\
	&e_4 = \frac{1}{8} G_2 (1 + \sqrt{1 + \frac{4 (2 - G_1) (2 - G_1 - G_2) \sin^2(2 \theta)}{G_2^2}}).
\end{align}
\end{small}
Similarly, $e_1$, $e_2$, $e_4 > 0$ and $e_3 < 0$, which implies that the post-measurement state is always entangled.

\subsection{Bilateral scenario}\label{Appdendix-A.2}

In the bilateral sequential scenario using weak measurement strategy, the partial transpose of Alice$_1$ and Bob$_1$'s post-measurement density matrix has four eigenvalues,
\begin{widetext}
	\begin{align}
		e_1 =& \frac{1}{16} \Big( 6 - G_1^2 + 2F_1 - \sqrt{20 + 12F_1 + 2 G_1^2 (-8 + 3 F_1) + (12 + 20F_1- 6 G_1^2 F_1) \cos(4 \theta) + 9 G_1^4 \sin^2(2 \theta)} \Big); \nonumber\\
		e_2 =& \frac{1}{16} \Big( 6 - G_1^2 + 2 F_1 + \sqrt{20 + 12F_1+ 2 G_1^2 (-8 + 3F_1 ) + (12 + 20F_1 - 6 G_1^2F_1) \cos(4 \theta) + 9 G_1^4 \sin^2(2 \theta)} \Big); \nonumber\\
		e_3 =& \frac{1}{16} \left( 2 + G_1^2 - 2 F_1 - \sqrt{25 G_1^4 + 8 (5 + 3 F_1) - 4 G_1^2 (16 + 5F_1)} \sin^2(2 \theta) \right); \nonumber\\
		e_4 =& \frac{1}{16} \left( 2 + G_1^2 - 2F_1 + \sqrt{25 G_1^4 + 8 (5 + 3 F_1) - 4 G_1^2 (16 + 5F_1)} \sin^2(2 \theta) \right).
	\end{align}
\end{widetext}
It is shown that \( e_1, e_2, e_4 >0 \), while \( e_3 <0 \) when 
\( \theta > \frac{1}{2} \arcsin\left(\frac{\sqrt{8 + G_1^4 - 8 F_1 - 4 G_1^2F_1}}{\sqrt{40 - 64 G_1^2 + 25 G_1^4 + 24F_1 - 20 G_1^2F_1 }}\right) \).

When $G_1=G_3=1$ and $\theta=\frac{\pi}{4}$, four eigenvalues are
\begin{small}
\begin{align}\label{ppt_twoside_weak}
		&f_1 = \frac{1}{16} \left(4 +F_4+F_2(1+2F_4)\right);
		&f_2 = \frac{1}{16} \left(6 + F_2 +F_4\right);\nonumber\\
		&f_3 = \frac{1}{16} \left(2-F_4 -F_2(1+2F_4) \right);
		&f_4 = \frac{1}{16} \left(4 - F_2 - F_4 \right).
\end{align}
\end{small}
We prove that \( f_1, f_2,f_4>0 \), whereas, $f_3<0$ only when $F_2<\frac{2-F_4}{1+2F_4}$, with \( F_{2},F_{_4} \in (0, \frac{1}{3}) \). 


For the PPM strategy with all information gain parameters equal, the four eigenvalues of the partial transpose of the intermediate post-measurement density matrix are
\begin{widetext}
\begin{small}
	\begin{align}
		u_1 &= \frac{1}{32} \left(16 + 2 (-4 + G_1) G_1 - \sqrt{2} \sqrt{64 + G_1 (-64 + G_1 (32 + 3 G_1 (-8 + 3 G_1))) + (64 - 64 G_1 + 24 G_1^3 - 9 G_1^4) \cos(4 \theta)}\right);& \nonumber\\
		u_2 &= \frac{1}{32} \left(16 + 2 (-4 + G_1) G_1 + \sqrt{2} \sqrt{64 + G_1 (-64 + G_1 (32 + 3 G_1 (-8 + 3 G_1))) + (64 - 64 G_1 + 24 G_1^3 - 9 G_1^4) \cos(4 \theta)}\right); &\nonumber\\
		u_3 &= \frac{1}{16} \left((4 - G_1) G_1 - \sqrt{(8 + G_1 (-12 + 5 G_1))^2 \sin^2(2 \theta)}\right);
		u_4 = \frac{1}{16} \left((4 - G_1) G_1 + \sqrt{(8 + G_1 (-12 + 5 G_1))^2 \sin^2(2 \theta)}\right).&
	\end{align}
\end{small}
\end{widetext}
 When \( \theta > \frac{1}{2} \arcsin\left(\frac{4 G_1 - G_1^2}{8 - 12 G_1 + 5 G_1^2}\right) \), the eigenvalues \( u_1, u_2 , u_4 >0\), while \( u_3<0 \). 



For the asymmetric PPM strategy with $\theta = \frac{\pi}{4}$ and $G_1 = G_3 = 1$, the eigenvalues coincide with those yielded by the weak measurement strategy (Eq.~\ref{ppt_twoside_weak}), leading to the same constraint $F_2<\frac{2-F_4}{1+2F_4}$ for \( F_{2},F_{_4} \in (0, \frac{1}{3}) \). However, the trade-off between information gain and state disturbance differs between the two strategies.

}

\bibliographystyle{apsrev4-1}
\bibliography{ref}

\end{document}